\documentclass[english,aps,prx,twocolumn,superscriptaddress,longbibliography]{revtex4-2}

\usepackage{mathrsfs}
\usepackage{upgreek}
 
\usepackage[symbol]{footmisc}
\RequirePackage{amssymb}
\RequirePackage{amsmath}
\RequirePackage{gensymb}
\RequirePackage{graphicx}
\RequirePackage{natbib}
\RequirePackage{nicefrac}
\RequirePackage{multirow}
\RequirePackage{epstopdf}
\RequirePackage{float}
\RequirePackage{physics}
\RequirePackage{braket}
\RequirePackage[linktocpage=true, colorlinks, allcolors=blue]{hyperref}
\RequirePackage{color} 
\RequirePackage{epigraph}
\RequirePackage{breqn}
\RequirePackage{bbold}
\RequirePackage[utf8]{inputenc}
\RequirePackage{newunicodechar}

\newboolean{ShowComments}
\setboolean{ShowComments}{true}  % change to true to show remaining colorful author comments. 
\ifthenelse{\boolean{ShowComments}}%
{
	\newcommand{\ColorComment}[3]{%
		{\colorbox{#1}{\color{white}   \textsf{\textbf{#2}}} \textcolor{#1}{#3}}}%  Colorful box, initials, phrase 
	\newcommand{\nyacite}[1]{[#1]}% not yet a cite
}%
{
	\newcommand{\ColorComment}[3]{}%  Do nothing at all
	\newcommand{\nyacite}[1]{}% not yet a cite -- do nothing
}%

\definecolor{jpcolor}{rgb}{0,0,1}
\definecolor{amcolor}{rgb}{0.2,0.5,0.4}

\definecolor{new}{rgb}{.38,.6,.38}
\definecolor{old}{rgb}{1,0,0}

\newcommand{\sixspace}{$\;\;\;\;\;\;$}

\newcommand{\princeton}{\affiliation{Department of Physics, Princeton University, Princeton, New Jersey~08544, USA}}

\newcommand{\NIST}{\affiliation{Joint Center for Quantum Information and Computer Science, NIST/University of Maryland, College Park, Maryland~20742, USA}}

\newcommand{\HRL}{\affiliation{HRL Laboratories, LLC, 3011 Malibu Canyon Road, Malibu, California 90265, USA}}

\begin{document}
\title{High fidelity state preparation, quantum control, and readout of an isotopically enriched silicon spin qubit}

\author{A. R. Mills}
\princeton
\author{C. R. Guinn}  
\princeton
\author{M. M. Feldman} 
\princeton
\author{A. J. Sigillito}  
\altaffiliation{Department of Electrical and Systems Engineering, University of Pennsylvania, Philadelphia, Pennsylvania~19104, USA}
\princeton
\author{M. J. Gullans}  
\NIST
\author{M. Rakher} 
\HRL
\author{J. Kerckhoff}   
\HRL
\author{C. A. C. Jackson} 
\HRL
\author{J. R. Petta}
\princeton

\begin{abstract}
Quantum systems must be prepared, controlled, and measured with high fidelity in order to perform complex quantum algorithms. Control fidelities have greatly improved in silicon spin qubits, but state preparation and readout fidelities have generally been poor. By operating with low electron temperatures and employing high-bandwidth cryogenic amplifiers, we demonstrate single qubit readout visibilities $>$99\%, exceeding the threshold for quantum error correction. In the same device, we achieve average single qubit control fidelities $>$99.95\%. Our results show that silicon spin qubits can be operated with high overall operation fidelity.
\end{abstract}

\maketitle

Backed by the highly successful semiconductor industry, the silicon spin qubit platform provides the potential to scale to large system sizes and integrate the classical control circuitry necessary for advanced operation protocols \cite{Loss1998,Ha2021,Xue_Intel_2021,BurkardRMP}. Since the earliest demonstrations of spin-qubit logic in GaAs \cite{Petta2005,Koppens2006}, a migration to isotopically enriched silicon \cite{Tyryshkin2012} combined with improvements in Si/SiGe heterostructure growth and device designs \cite{Deelman2016,Zajac2015} have led to a recent surge of demonstrations of single- and two-qubit gates with fidelities above 99\% \cite{Xue2022,Noiri2022,Mills2022}.

In order for a qubit platform to be a serious contender for quantum information processing it must be able to demonstrate all  of the DiVincenzo criteria for quantum computing with high fidelity. While single- and two-qubit gates implemented in Si have made steady progress, state preparation and measurement (SPAM) fidelities have generally been well below 90\%, with a few recent exceptions \cite{Andrews2019,Blumoff2021,Mills2022,Philips_6QB_arxiv}. To implement quantum error correction and realize fault tolerant operation the total logical error rate, which includes SPAM, must be kept low $\lesssim$ 2\% \cite{Fowler2012}.  

Depending on the qubit encoding, there are various protocols for initializing and reading out spin qubits \cite{Blumoff2021,Borjans2020}, and a combination of techniques will likely be required for larger spin qubit systems. Currently, readout in singlet-triplet and exchange-only qubits is performed using Pauli spin blockade \cite{Petta2005,Medford2013}, whereas single-spin qubits typically use Elzerman readout. The Elzerman approach utilizes state dependent tunneling to prepare and measure spin qubits \cite{Elzerman2004}. Protocols taking advantage of enhanced spin-charge relaxation in double quantum dots (DQDs) can be used to accelerate spin initialization or implement spin initialization in isolated DQDs that are not strongly tunnel coupled to leads \cite{Srinivasa2013,Xue2019}. Finally, there are a variety of schemes to improve the signal-to-noise-ratio (SNR) and measurement bandwidth through the use of cryogenic amplifiers \cite{Vink2007}, RF reflectometry \cite{Reilly2007fast}, and latched charge and spin readout techniques \cite{Petersson2010,Studenikin2012,Harvey2018}.

In this Letter, we demonstrate a readout visibility greater than 99\% and average single qubit gate fidelities above 99.95\% in a single spin Loss-DiVincenzo (LD) qubit. Cryogenic amplifiers and circuit optimization allow for low noise, high bandwidth (1 MHz) charge sensing with a charge detection SNR $>$ 12. This high SNR, when combined with optimized spin readout parameters, enables high visibility Elzerman spin state readout \cite{Elzerman2004}. High fidelity single spin rotations in the same spin qubit are achieved using electric dipole spin resonance \cite{Tokura2006,Pioro2008}, as verified by interleaved randomized benchmarking (IRB) \cite{Magesan2011}. These results show that overall operation fidelities in Si spin qubits can exceed important thresholds for fault-tolerant operation. 

\begin{figure*}[t]
	\centering
	\includegraphics[width=2\columnwidth]{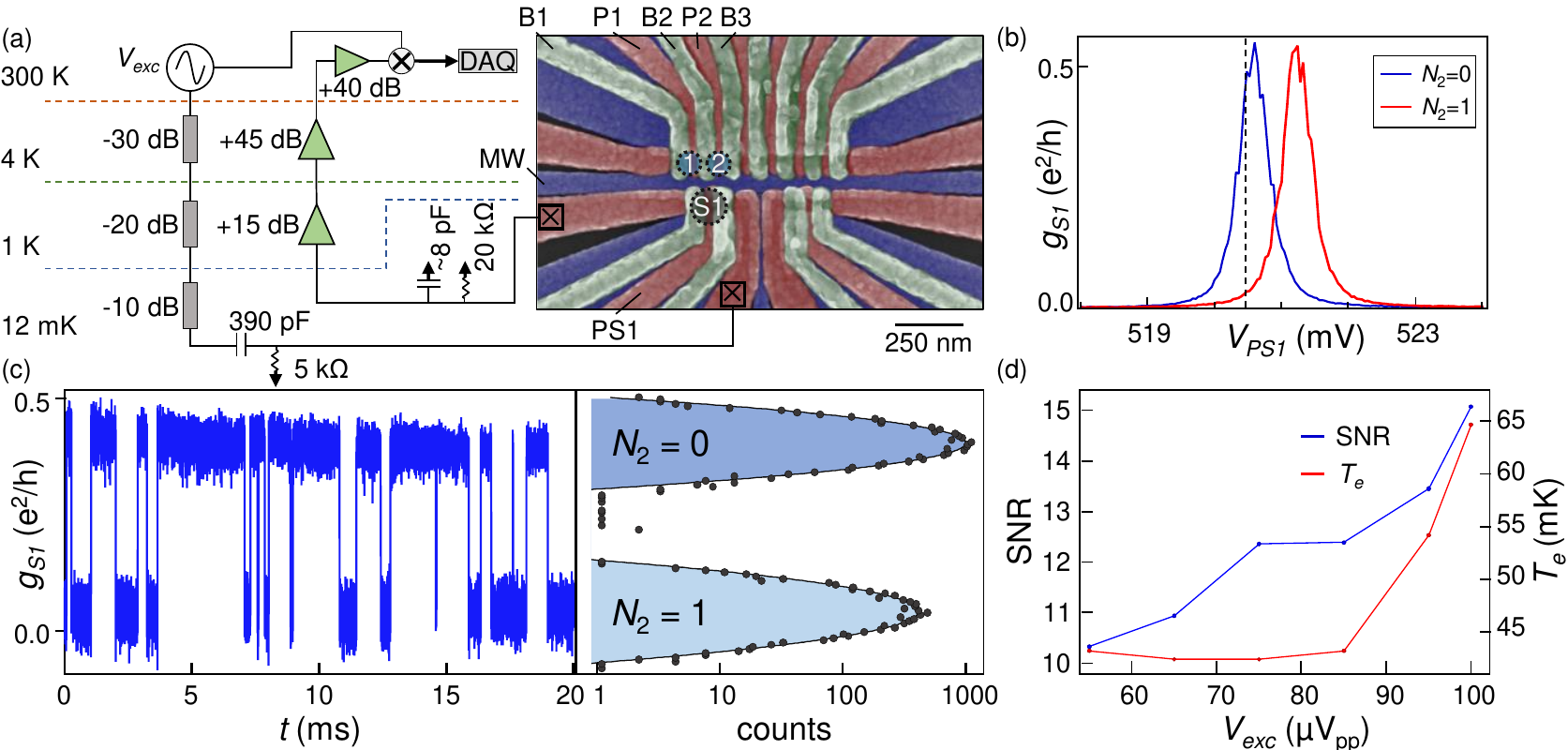} 
	\caption{High fidelity charge state determination. 
	(a) False-color scanning electron microscope image of the device showing the two quantum dots and the charge sensor ``S1'' utilized for spin-to-charge conversion. The illustration to the left shows the charge sensing circuit. An ac signal $V_{\rm exc}$ is highly attenuated before being ac-coupled to the sensor dot. Two HEMT amplifiers measure the voltage drop across a 20 k$\Omega$ resistor that is mounted to the sample holder. Parasitic capacitance before the first stage amplifier is indicated alongside the surface mounted resistor. The signal is further amplified at room temperature before demodulation and digitization (DAQ).	
	(b) A large shift of the sensor dot Coulomb blockade peak is evident when the occupancy of dot 2 changes from $N_2=0$ to $N_2=1$. The sensor bias point is indicated with a dashed line. 
	(c) Time-series of the charge sensor conductance $g_{\rm S1}$ sampled at 1 MS/s and with $V_{exc} = 85 \;\mu V_{pp}$ showing real-time tunneling events between $N_2=0$ to $N_2=1$ (left). Histogram of the time-series data used to extract the charge readout SNR (right). 
	(d) Charge readout SNR and electron temperature $T_e$ for dot 2 measured as a function of  $V_{\rm exc}$.
    }
	\label{fig:1}
\end{figure*}

The device consists of a Si/SiGe heterostructure with an isotopically purified $^{28}$Si (800 ppm residual $^{29}$Si) quantum well. Lithographically defined overlapping aluminum gate electrodes are used to define a linear array of 6 quantum dots with 2 proximal charge sensors \cite{Zajac2015,Zajac2016}. High fidelity state preparation, control, and measurement are demonstrated in a single LD qubit formed under gate P2 and a proximal charge detector S1 is used to read out the charge state of the quantum dot [Fig.~1(a)]. Microwaves are applied to the center MW gate to perform single qubit rotations using electric dipole spin resonance in the field gradient of a Co micromagnet \cite{Pioro2008}. Dot 1 is kept empty $(N_1 = 0)$ for these experiments and dot 2 is coupled to the reservoir via an accumulated channel on the right side of the device \cite{Zajac2016}. 

Spins are selectively prepared and measured using spin-to-charge conversion \cite{Elzerman2004}. To obtain a 99\% readout visibility both electrical detection and spin-to-charge conversion have to function with high fidelity. For high fidelity electrical detection, the measurement noise needs to be much lower than the charge sensing signal associated with the spin-dependent tunneling events. As we will demonstrate, robust charge sensing is feasible in Si spin qubit devices and is generally not the limiting factor in the overall readout visibility. On the other hand, the tunnel rates, magnetic field, readout bias point, and signal sampling rate must be carefully optimized. The requirements for spin-to-charge conversion with visibility exceeding 99\% are reviewed thoroughly by Keith \textit{et al.} \cite{Keith2019}. For each condition, the minimum requirement for achieving 99\% visibility is: 1) a large Zeeman splitting $E_Z$ relative to the electron temperature $T_e$, $E_Z \gtrsim 13k_BT_e$, 2) a fast tunnel out time $t^{\uparrow}_{out}$ for a spin-up electron relative to the spin relaxation time $T_1$, $T_1 \gtrsim 100t^{\uparrow}_{out}$, and 3) a fast sampling rate $\Gamma_s$ relative to the reload rate $1/t^{\downarrow}_{in}$, $\Gamma_s \gtrsim 12/t^{\downarrow}_{in}$. If any of these requirements are not met, 99\% visibility Elzerman spin readout is not possible \cite{Keith2019}. However, just barely meeting all of these requirements will also result in $<99\%$ visibility. In practice, these three conditions must be budgeted such that the combined infidelities are $<1\%$ as discussed below.

We first optimize charge state readout using the circuit shown in Fig. 1(a). A 1 MHz sine wave is applied to S1 and the drain current flows to ground through a 20 k$\Omega$ resistor. The voltage drop across the 20 k$\Omega$ resistor is amplified at the 1 Kelvin still plate (+15 dB gain) and 4 Kelvin plate (+45 dB gain) before reaching a room temperature amplifier \cite{Blumoff2021}. The signal is then demodulated and digitized. Before the first stage amplifier there is $\sim$8 pF of parasitic capacitance which limits the circuit bandwidth to $\sim$1 MHz. Figure 1(b) shows a Coulomb blockade peak in the charge sensor conductance $g_{\rm S1}$ as the sensor dot plunger gate voltage, $V_{\rm PS1}$, is swept. Changing the electron number in dot 2 from $N_2$ = 0 to $N_2$ = 1 shifts the Coulomb blockade peak by approximately its full width at half maximum. When biased on the side of a Coulomb blockade peak the sensor dot can easily detect  real-time tunneling events, as we now demonstrate.

High bandwidth charge detection is illustrated in Fig.~1(c), where we show a time-series of $g_{\rm S1}$ sampled at 1 MS/s with the chemical potential of dot 2 tuned close to the Fermi level of the reservoir. Real-time tunneling events between the $N_2$ = 0 and $N_2$ = 1 charge states are visible. The switching rate between these charge states is set by the tunnel coupling $\Gamma$ between the dot and the reservoir, tuned here to be slower than our measurement bandwidth. A histogram of these data are fit by a double-Gaussian curve with center positions $\mu_n$ and standard deviations $\sigma_n$. The charge readout SNR is set by the separation of the two Gaussians relative to their spread: SNR = $(\mu_2-\mu_1)/(\bar{\sigma})$ where we use $\bar{\sigma} = (\sigma_1+\sigma_2)/2$ to account for slightly different standard deviations in the double-Gaussian. 

Heating from the charge sensor is explored in Fig. 1(d), where we plot the SNR and electron temperature $T_e$ as a function of the peak-to-peak excitation voltage $V_{exc}$ at the sensor.  The SNR increases with $V_{exc}$ as expected, but for $V_{exc}$ $>$ 85 $\mu V_{pp}$ a steady increase of $T_e$ with $V_{exc}$ is observed. We therefore operate with $V_{exc}$~$=$ 85~$\mu V_{pp}$, where the SNR $\approx$ 12.5 and $T_e$ $\approx$ 45 mK. The electron temperature is estimated by the broadening of the tunneling line width for the first electron dot-reservoir transition. These electron temperatures are significantly lower than values ($\sim 200$ mK) that have been reported in prior single-shot readout experiments with Si devices, allowing us to operate at lower fields while maintaining high measurement visibility \cite{Keith2019b,Morello2010}. We attribute our low electron temperature to proper thermalization of the device at the mixing chamber and the careful elimination of ground loops.

In theory, a charge sensing SNR = 12.5 yields a lower bound estimate of the charge state infidelity $1 - F_c \geq 3 \times 10^{-10}$ \cite{Blumoff2021}. Experimental non-idealities in the charge sensing signal, such as charge fluctuations in the device, can impact this infidelity. Regardless, the negligible charge state infidelity implies that the overall readout performance will be limited by the spin-to-charge conversion process. We now explore the parameters that must be optimized for high-fidelity spin-to-charge conversion.

\begin{figure}[t]
	\centering
	\includegraphics[width=\columnwidth]{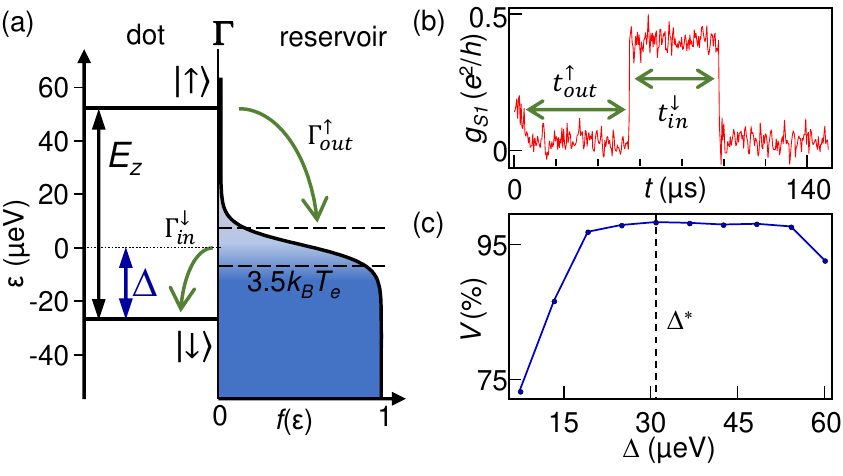} 
	\caption{Optimization of the physical readout parameters. (a) Quantum dot energy levels tunnel coupled to a thermally broadened Fermi reservoir. (b) Time-series of a typical spin-up detection event showing the measured tunneling times $t^{\downarrow}_{in}(t^{\uparrow}_{out})$ that are used to estimate the rates $\Gamma^{\downarrow}_{in}(\Gamma^{\uparrow}_{out})$.
	(c) Measurement visibility $V$ as a function of Fermi offset $\Delta$. The offset $\Delta^*$ that maximizes $V$ is then used in subsequent measurements.}
	\label{fig:2}
\end{figure}

Figure 2(a) illustrates the process of spin-to-charge conversion for a spin-up electron. During the readout phase of an experiment, the energy of the spin-up and spin-down states of the quantum dot electron are set to straddle the Fermi level of the reservoir. The spin-up electron tunnels off the dot on a time-scale set by 1/$\Gamma^{\uparrow}_{out}$ and is then replaced by a spin-down electron that tunnels into the dot on a time-scale set by 1/$\Gamma^{\downarrow}_{in}$ \cite{Elzerman2004}. This sequence of events creates a detectable `spin bump' in the charge sensing signal, Figure 2(b). $\Gamma^{\uparrow}_{out}$ should be fast compared with the spin relaxation rate $1/T_1$, but the reload rate $\Gamma^{\downarrow}_{in}$ must be slow enough to allow the short change in electron occupancy to be detectable given the finite bandwidth of the measurement circuit. The overall tunnel rate $\Gamma$ to the reservoir accumulated to the right of dot 2 is set by the barrier gate voltage $V_{B3}$, while the ratio of the tunnel rates $\Gamma^{\uparrow}_{out}/\Gamma^{\downarrow}_{in}$ can be tuned by adjusting the parameter $\Delta$, which is the energy difference between the spin-down state and the Fermi level of the reservoir.

Thermally activated tunneling events can significantly impact the performance of energy dependent state preparation and measurement. For example, the probability of a spin-down electron tunneling into an unoccupied state of the Fermi sea can be non-negligible. Moreover, state preparation errors can occur when spin-up states tunnel onto the dot during the reloading period. To achieve high-fidelity spin-to-charge conversion, the Zeeman splitting $E_Z$ must be much larger than $T_e$. Increasing the Zeeman splitting can suppress thermal errors, but $E_Z$ is constrained by enhanced excited state relaxation at higher magnetic fields \cite{Borjans2019} and also by practical constraints on microwave signal generation and delivery to the device. Here we operate at $B_{ext} = 410$ mT, with $E_Z = 19.105$~GHz (79 $\mu$eV) and $T_1 = 31.5$~ms.

In addition to optimizing the ratio $E_z/k_BT_e$, the parameter $\Delta$ must also be carefully tuned to limit thermally activated tunneling events. Thermal excitation of a spin-down electron can be suppressed by increasing $\Delta$, but the tradeoff is that $\Gamma^{\uparrow}_{out}$ decreases and $\Gamma^{\downarrow}_{in}$ increases. The decrease in $\Gamma^{\uparrow}_{out}$ slows down the spin-to-charge conversion process resulting in $T_1$ relaxation errors, while the increase in $\Gamma^{\downarrow}_{in}$ makes the charge hopping events shorter and harder to detect with our 1 MHz measurement bandwidth. Therefore, the optimal $\Delta$ is large enough to suppress thermal errors and small enough to maximize the ratio $\Gamma^{\uparrow}_{out}/\Gamma^{\downarrow}_{in}$. The rates $\Gamma^{\uparrow(\downarrow)}_{out(in)}$ are extracted by binning the tunneling times from many single shot traces [one is shown in Fig. 2(b)] into a histogram and fitting to an exponential decay. To optimize $\Delta$ we perform 10,000 measurements interleaving spin-up and spin-down prepared states and measure the visibility $V = F_{\uparrow} + F_{\downarrow} - 1$. Figure 2(c) shows the measurement visibility $V$ as a function of $\Delta$, with the optimal value $\Delta^* \approx 30 \;\mu$eV resulting in $\Gamma^{\uparrow}_{out} \approx \Gamma^{\downarrow}_{in} \approx 20$ kHz. To counteract slow drift in the device during long quantum control sequences we periodically recalibrate to maintain the optimal $\Delta^*$.

With the physical parameters described above optimized for high fidelity spin-to-charge conversion, we turn to the optimization of data acquisition parameters, namely the conductance threshold $g_{thr}$ and duration of the readout window $t_R$. A spin-up state is registered whenever $g_{\rm S1}$ exceeds $g_{thr}$ within the analysis time-window $t_R$ [Fig. 3(a)]. If $g_{thr}$ is set too low, then background noise can lead to false positives and reduce the spin-down fidelity $F_{\downarrow}$. On the other hand, if $g_{thr}$ is set too high then we begin to miss the short, near bandwidth-limited hopping events that may not reach full amplitude, resulting in a reduced spin-up fidelity $F_{\uparrow}$. The time $t_R$ should be long enough to catch all hopping events from spin-to-charge conversion. However, if $t_R$ greatly exceeds the characteristic tunneling time $t^{\uparrow}_{out}$ then more thermal errors will occur, limiting $F_{\downarrow}$. In larger arrays with sequential readout steps, the readout time will also need to be balanced against $T_1$ decay in the subsequent qubits adding an additional constraint on $t_R$ and the tunneling rates \cite{Mills2022}.

\begin{figure}[t]
	\centering
	\includegraphics[width=\columnwidth]{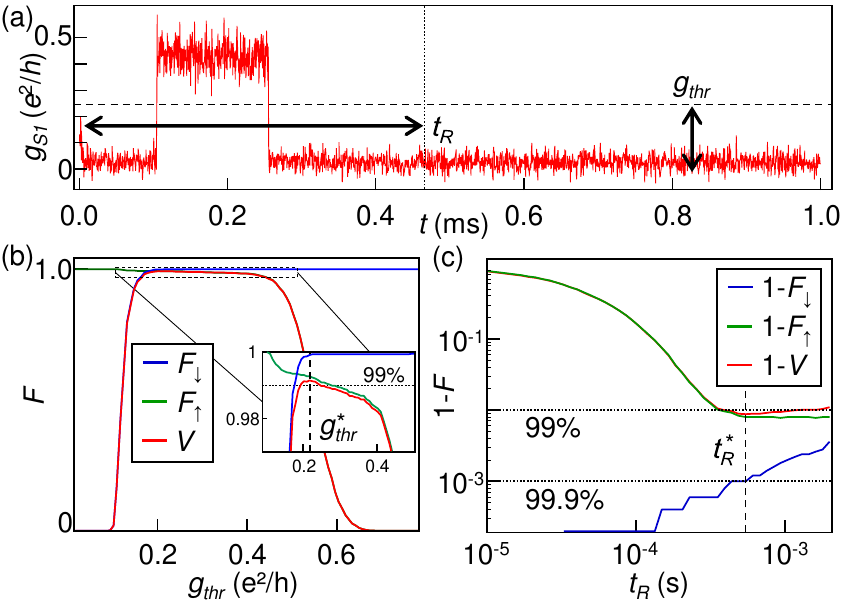} 
	\caption{Optimization of the software readout parameters.
	(a) Typical time-series with the conductance threshold $g_{thr}$ and end-of-read window $t_R$ indicated.
	(b) Measurement fidelity for spin-up(spin-down) states $F_{\uparrow}$($F_{\downarrow}$) and the overall measurement visibility $V$ as a function of $g_{thr}$. The optimum threshold $g_{thr}^*$ is indicated in the inset with a vertical dashed line. 
	(c) Measurement infidelities plotted on a logarithmic scale as a function of $t_R$. The optimum end of read window $t_{R}^*$ is indicated with a vertical dashed line. With both parameters optimized we achieve $F_{\downarrow} = 99.86\%$, $F_{\uparrow} = 99.26\%$, and $V = 99.12\%$.}
	\label{fig:3}
\end{figure}

We perform another 10,000 measurements interleaving spin-up and spin-down prepared states using the optimized $\Delta^*$ from above in order to demonstrate the optimization of $g_{thr}$ and $t_R$. In Fig.~3(b), the measurement fidelities $F_{\uparrow(\downarrow)}$ and visibility $V = F_{\uparrow} + F_{\downarrow} - 1$ are plotted as a function of $g_{thr}$ showing the optimum $g_{thr}^* = 0.22 \: e^2/h$. The slight negative slope at the top of the visibility curve indicates we are near bandwidth limitations as the very short hopping events are effectively low-pass filtered and unable to reach full amplitude. Figure 3(c) shows the measurement infidelities $1-F$ as a function of $t_R$. The highest readout visibility is obtained with $t_R^* = 670 \: \mu s$. Sampling beyond this time slowly increases $1-F_{\downarrow}$ due to thermal tunneling events, which reduces the overall visibility. These fidelity estimates inherently include errors from both state preparation and measurement.

With all of the parameters optimized we are able to achieve detection fidelities $>$99\%, with $F_{\downarrow} = 99.86\% \pm 0.05\%$ and $F_{\uparrow} = 99.26\% \pm 0.12\%$ yielding an average measurement fidelity $F_M = 99.56\%$. Statistical error is estimated as one standard deviation from binomial sampling. The spin-up fidelity levels off at 99.26\% due to a loss of spin information from $T_1$ relaxation ($\sim 0.2\%$) and missed spin bumps caused by bandwidth limitations in the amplification chain ($\sim 0.5\%)$. Relaxation errors are calculated using the characteristic tunneling time $t^{\uparrow}_{out} = 50 \;\mu s$ plus a 10 $\mu s$ readout settling time, and $T_1$. The probability of missing a spin bump $P_{miss} = 1 - \dfrac{(1-e^{(R^{\uparrow}_s - R^{\downarrow}_s)/2})R_s}{(1-e^{R^{\uparrow}_s/2})(R^{\uparrow}_s-R^{\downarrow}_s)}$, where $R^{\uparrow}_s = t_s/t^{\uparrow}_{out}$, $R^{\downarrow}_s = t_s/t^{\downarrow}_{in}$, and $t_s$ = 1 $\mu$s is the sampling time \cite{Keith2019}. We estimate an additional $P^{\downarrow}_{out} = 0.06\%$ error due to thermal excitations using $P^{\downarrow}_{out} = e^{-t^{\downarrow}_{out}/t_R}$, where $t^{\downarrow}_{out}$ is estimated from $t^{\downarrow}_{out} = t^{\downarrow}_{in} e^{E_z/2k_BT_e}$ \cite{Keith2019}. Tunneling out of the spin-down state reduces the spin-down fidelity and the remaining $\sim 0.1\%$ of error is likely due to slow drift of $\Delta$ during measurement. 

Past experiments demonstrating high fidelity control of LD spin qubits were generally limited to $V$ $\approx$ 70--80\% \cite{Yoneda2018,Huang2019,Xue2019}. Recent experiments on a six qubit device achieved $V$ = 93.5 --  98\% \cite{Philips_6QB_arxiv}. Here we demonstrate the integrated high performance of our device. Figure 4(a) shows the spin-up probability $P_\uparrow$ plotted as a function of the frequency detuning $\Delta f$ from resonance (19.105 GHz) and the microwave burst length $\tau_R$. Rabi oscillations are obtained when driving on resonance [Fig.~4(b)]. We rigorously verify high gate and SPAM fidelities using gate set tomography (GST) protocols for single qubit gates ($I,X,Y$) \cite{Nielsen2021} where an $X(Y)$ gate is a $\pi/2$ rotation performed about the $X(Y)$ axis and $I$ is performed by idling the qubit for the same amount of time as the $X(Y)$ rotations. GST yields a state preparation fidelity $\rho_0 = 99.76\% \pm 0.04\%$ and a measurement fidelity $M = 99.35\% \pm 0.1\%$, which is consistent with the Fig.~3 data. The average single qubit gate fidelities extracted from GST are 99.956\% $\pm 0.002\%$. The gate fidelity is primarily limited by incoherent noise caused by qubit dephasing ($T_2^*=3.2\: \mu s,\, T_2^H=139\: \mu s$ measured using Ramsey and Hahn echo pulse sequences). Due to the modest $T_2^*$, the idling fidelity $F_I = 99.43\% \pm 0.036\%$ is significantly lower than fidelities obtained during driven evolution.

\begin{figure}[tbp]
	\centering
	\includegraphics[width=\columnwidth]{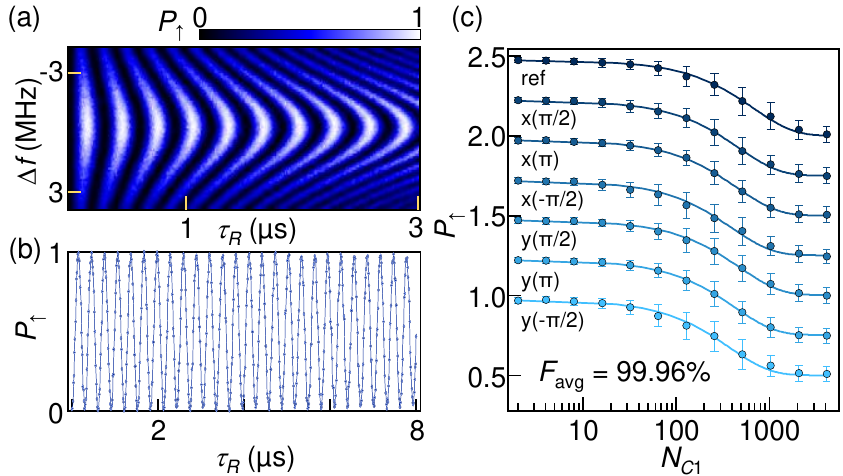}
	\caption{Single qubit randomized benchmarking. (a) $P_{\uparrow}$ as a function the frequency detuning $\Delta f$ from resonance (19.105 GHz) and $\tau_R$.
	(b) High visibility Rabi oscillations from driving the qubit on resonance.
	(c) Return probability $P_{\uparrow}$ as a function of $N_{C1}$ for each IRB curve (including the reference) showing high visibility, full saturation, and average gate fidelity 99.96\%. The curves are offset by 0.25 for clarity.
	}
	\label{fig:RB}
\end{figure}

Finally, we perform IRB to estimate fidelities for the interleaved gates $(X,X^2,-X,Y,Y^2,-Y)$ \cite{Magesan2012}, where a $X^2(Y^2)$ gate is a full $\pi$ rotation about the respective axis. To obtain reliable results we utilize $k = 200$ unique sequences per point, with 100 averages. We chose $k = 200$ to obtain a rigorous gate error estimate as $1/f$-noise dominated systems see diminishing returns on the accuracy of this estimate when increasing $k$ beyond $\sim$100 \cite{Epstein2014}. Sequence lengths of up to  $N_{C1} = 4096$ Clifford operations are employed to achieve full saturation of the sequence fidelity curves [Fig. 4(c)]. The error bars on each point are the standard deviation of the fidelities for the 200 unique sequences at each point. The average gate fidelities are shown in Table 1 with error bars determined using bootstrapping, a technique to randomly resample within the full data set to build statistics \cite{Barends2014}.  Retuning routines are implemented at $\sim30$ min. intervals during these long measurements ($\sim$ 14 hrs.)  to correct for readout and qubit frequency drift. Moreover, to reduce heating at the device, the charge sensor excitation is turned off during qubit manipulation.

\begin{table}
\begin{tabular}{ |p{1cm}||p{2.9cm}|p{1.5cm}||p{2.8cm}| }
 \hline
 \multicolumn{2}{|c|}{IRB Fidelities} &
 \multicolumn{2}{|c|}{GST Fidelities} \\
 \hline
 Gate & Fidelity & Operation & Fidelity \\
 \hline
 $\:\;\;X$  & 99.969\%  $\pm 0.004\%$   & \sixspace$\rho_0$  & 99.76\% $\pm 0.04\%$ \\
 $\:\;\;X^2$& 99.964\%  $\pm 0.003\%$   & \sixspace$M$       & 99.35\% $\pm 0.1\%$\\
 $-X$       & 99.949\%  $\pm 0.005\%$   & \sixspace$I$       & 99.43\% $\pm 0.036\%$ \\
 $\:\;\;Y$  & 99.973\%  $\pm 0.004\%$   & \sixspace$X$       & 99.958\% $\pm 0.002\%$\\
 $\:\;\;Y^2$& 99.961\%  $\pm 0.004\%$   & \sixspace$Y$       & 99.954\% $\pm 0.002\%$\\
 $-Y$       & 99.937\%  $\pm 0.005\%$   &           & \\
 \hline
\end{tabular}
\caption{Interleaved randomized benchmarking gate fidelities. Average Clifford fidelity for the reference is 99.85\% corresponding to an average gate fidelity of 99.96\%, in agreement with extracted gate fidelities. On the right, GST results for SPAM, identity, and gate operations X and Y with average gate operation fidelity 99.956\%}
\end{table}
 
In conclusion, our measurements show that Si spin qubits can be operated reliably with all-around high performance metrics. Optimal state preparation and measurement requires careful balancing of physical constraints with hardware constraints to minimize the loss of spin information due to spin relaxation and a finite 1 MHz measurement bandwidth. We are able to achieve measurement fidelities exceeding 99\%, as verified through the analysis of single-shot readout traces and GST. GST and IRB are implemented to demonstrate average single qubit gate fidelities exceeding 99.95\% under the same operating conditions. Elzerman readout of larger LD spin qubit arrays \cite{Zajac2016,Sigillito2019site,Mills2022,Philips_6QB_arxiv} will require a reduction of the measurement time relative to the spin relaxation time. Furthermore, faster readout protocols will be necessary to fully unlock the potential of feedback-based error correction protocols \cite{Terhal2015}.

\begin{acknowledgements}
Supported by Army Research Office grant W911NF-15-1-0149 and DARPA grant D18AC0025. Devices were fabricated in the Princeton University Quantum Device Nanofabrication Laboratory, which is managed by the department of physics. The authors acknowledge the use of Princeton's Imaging and Analysis Center, which is partially supported by the Princeton Center for Complex Materials, a National Science Foundation MRSEC program (DMR-2011750).
\end{acknowledgements}

\bibliography{PettaLab_Refs}

\end{document}